\documentclass[aps,prb,twocolumn,showpacs]{revtex4}
\usepackage{graphicx}
\usepackage{amssymb}
\usepackage{mathrsfs}
\newcommand{\bea}[1]{\begin{eqnarray}\label{#1}}
\newcommand{\eea}{\end{eqnarray}}

\begin{document}
  \title{Ultrafast hot carrier dynamics of ZrTe$_5$ from time-resolved optical reflectivity}
  \author{Xiu Zhang$^{1}$, Hai-Ying Song$^{1}$, X. C. Nie$^{1}$, Shi-Bing Liu$^{1}$, Yang Wang$^{1}$, Cong-Ying Jiang$^{1}$, Shi-Zhong Zhao$^{1}$}
 \author{Genfu Chen$^{2}$}
  \author{Jian-Qiao Meng$^{3}$, Yu-Xia Duan$^{4}$}
 \author{H. Y. Liu$^{1}$}
   \email[Corresponding author: ]{haiyun.liu@bjut.edu.cn}

\affiliation{
\\$^{1}$Strong-field and Ultrafast Photonics Lab, Institute of Laser Engineering,
Beijing University of Technology, Beijing 100124, P. R. China
\\$^{2}$Beijing National Laboratory for Condensed Matter Physics, Institute of Physics, Chinese Academy of Sciences, Beijing 100190, China
\\$^{3}$Hunan Key Laboratory of Super-microstructure and Ultrafast Process, School of Physics and Electronics,
Central South University, Changsha, Hunan 410083, P. R. China
\\$^{4}$School of Physics and Electronics, Central South University, Changsha, Hunan 410083, P. R. China
}

\begin{abstract}
 We investigate the hot carrier dynamics of ZrTe$_5$ by ultrafast time-resolved optical reflectivity. Our results reveal a phonon-mediated across-gap recombination, consistent with its temperature-dependent gap nature as observed previously by photoemission. In addition, two distinct relaxations with a kink feature right after initial photoexcitation are well resolved, suggesting the complexity of the electron thermalization process. Our findings indicate that correlated many-body effects play important role for the transient dynamics of ZrTe$_5$.
\end{abstract}

\date{\today}
\maketitle
\section{INTRODUCTION}
Transition metal pentatellurid ZrTe$_{5}$ has recently triggered considerable interests for its unique electronic properties. It is predicted to be on the boundary between strong and weak topological insulators depending on its lattice parameter\cite{Weng-HM,TI-phase}. Angle-resolved photoemission spectroscopy (ARPES) measurements have evidenced its temperature-dependent electronic structure from a p-type semimetal to a semiconductor and to a n-type semimetal, associated with a bandgap variation from $\sim$ 30 to $\sim$ 20 meV\cite{Zhou-XJ,ZX-ARPES-2017}, providing clear interpretation for the strong resistivity peak and the sign reversal of Hall effect near 135 K\cite{anomalies-resistivity1,anomalies-resistivity2,anomalies-resistivity3}. ZrTe$_{5}$ hosts both 1D chain and 2D layer structures\cite{Zhou-XJ, 1986-FH-SSC} with weak interlayer coupling comparable to that of graphite\cite{Weng-HM}. In addition, it has high in-plane carrier mobility comparable to that of graphene\cite{2017-NLWang-PNAS} and strong electron-electron (e-e) interaction\cite{2016-SR-Hall}. These remind us with some low-dimensional quantum-confined systems with enhanced Coulomb interaction and carrier-carrier scattering, such as graphene\cite{graphene2} and semiconductor nanocrystals\cite{nanocrystal1-2008,nanocrystal2-2007,2005-Ncrystal-200fs,nanocrystal3-2006}. At this stage, comprehensive ultrafast time-resolved studies are required to unveil many-body interactions in ZrTe$_{5}$.

Time-resolved optical spectroscopy is an effective way of tracking photoexcited hot carrier dynamics in ultrafast timescales from few femtoseconds (fs) to nanoseconds (ns). In general, for gapped systems under weak excitation (too weak to break the ordered state), hot carriers after initial photoexcitation (at $t_1$) relax to equilibrium in two steps as shown in Fig. 1a: (i) $t_1 < t < t_2$, hot carriers rapidly lose energy by fast e-e scattering and electron-phonon (e-ph) scattering, and accumulate above the gap around the Fermi level ($E_F$), referring to the thermalization process; (ii) $t > t_2$, across-gap recombination occurs by the emission of gap-frequency phonons, resulting in the gap- and phonon- sensitive equilibration process. In Fig. 1a we use the sketch of band structure at $\Gamma$ in ZrTe$_5$, where most of the electronic spectral weight was found around $E_F$ by ARPES\cite{Zhou-XJ}, corresponding to the highest excitation probability. The optical probe is dedicated to measure the relaxation of the dielectric constant resulting from the transient electronic distribution, for example, in the equilibration process (ii) the density of hot carriers is proportional to spectroscopy changes in transmission/reflectivity (Fig. 1b)\cite{1999-JD-CDWprl,2001-JD-PP}. Therefore time-resolved optical spectroscopy allows direct investigation of carrier dynamics, gap properties and many-body effects in many materials, such as graphene\cite{Pp-graphene1}, superconductors\cite{Pp-SC1,Pp-SC2,tr-prl-hotcarrier,2018-nie-2212}, semiconductors \cite{Pp-semiconductor2} and charge density wave materials\cite{Pp-CDW1,Pp-CDW2,Pp-CDW3}. In most materials, the thermalization process (i) has a very short time constant within few tens of fs\cite{1999-JD-CDWprl}, resulting in a sharp signal rising edge that is comparable to the duration of pump and probe pulses. However, in some low-dimensional quantum-confined systems with enhanced Coulomb coupling such as graphene\cite{graphene2} and semiconductor nanocrystals\cite{2005-Ncrystal-200fs}, e-e scattering has a time constant up to 200 fs, offering a opportunity to directly study related effects such as quasiparticle multiplication and impact excitation.

 \begin{figure}                                                                                                                                             \includegraphics[width=1\columnwidth,angle=0]{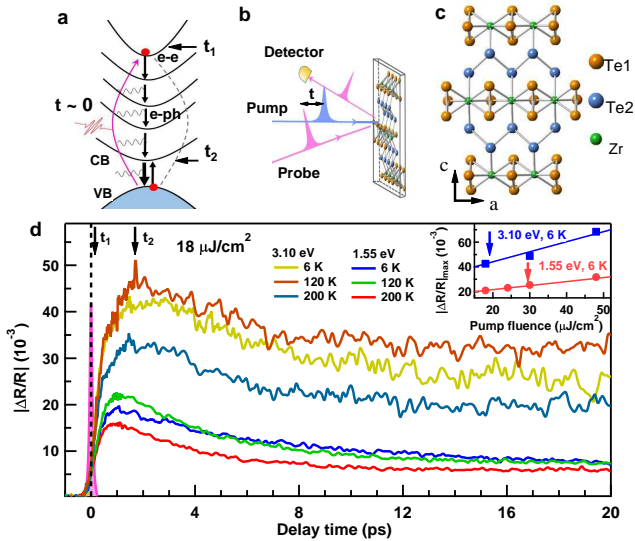}
   \caption{Time-resolved reflectivity changes of ZrTe$_{5}$. (a) Photoexcitation and relaxation processes. The initial photoexcitation time is marked as $t_1$. The BCB buildup time is denoted as $t_2$. Hot electrons from high-lying conduction band ($t_1$) lose energy and accumulate to the bottom of the low conduction band ($t_2$). The dashed line illustrates the impact excitation for e-e scattering, in which a photoexcited electron collides with an electron in the valence band. (b) Schematic of ultrafast time-resolved optical reflectivity.(c) The a-c plane of the ZrTe$_{5}$ crystal sheet, formed by ZrTe$_{3}$ chains along the a-axis and linked via zig-zag chains along the c-axis. The bulk crystal is constructed by stacking the ZrTe$_{5}$ sheets along the b-axis (perpendicular to a-c plane) with very weak interlayer coupling. The blue and orange spheres represent Te atoms, while Zr atoms are marked as the green ones. (d) Representative reflectivity changes at 6, 135 and 300 K, using pump pulses of 3.10 and 1.55 eV with pump fluence $F$ = 18 $\mu$$ J/cm^{2}$. The zero time delay is determined by the pink shaded curve, which is the cross-correlation of the pump and probe beams. The black arrow denotes the maximum of $|\Delta R/R|$, corresponding to the BCB buildup at $t_2$.}
\end{figure}

In this paper, we investigate the hot carrier dynamics in ZrTe$_{5}$ by ultrafast time-resolved optical reflectivity, by inter-band photoexcitation using two pump photon energies: 1.55 eV (800 nm) and 3.10 eV (400 nm). Our results reveal a relaxation bottleneck in the equilibration process where a phonon-mediated across-gap recombination occurs, consistent with the gap nature and phase transitions. In the thermalization process, we find a resolvable rising edge for the buildup dynamics of the bottom of the conduction band (BCB buildup) with a constant kink feature at 500 (fs), allowing us to investigate the complexity in the transient dynamics.

\section{METHOD}
In our experimental setup, the infrared pulses were generated by a mode-locked Ti:sapphire regenerative amplifier system working at 1 KHz repetition rate, centered at 800 nm. Both the pump and probe beam were focused onto the a-c plane of a freshly cleaved sample surface in vacuum (Fig. 1b), with spot sizes at $\sim$ 0.4 mm (pump) and $\sim$ 0.2 mm (probe) in diameter. The 400 nm pump was obtained by second harmonic generation using a nonlinear phase-matched barium borate (BBO) crystal (thickness 0.5 mm). The reflected probe signal was collected by a Si-based detector and a lock-in amplifier. The time delay was realized by scanning the delay time between pump and probe pulses, using a motorized delay line. The sample was mounted on a cryostat with a temperature sensor embedded nearby, allowing a precise control of temperatures from 6 to 300 K. High quality ZrTe$_5$ single crystal was grown by chemical vapor transport technique with iodine methods\cite{chemical-vapor}. The crystal structure consists of ZrTe$_{3}$ sheets extend along the b axis (perpendicular to the a-c plane) forming a layered structure (Fig. 1c), and the adjacent two sheets is quite large and held together by weak interlayer van der Waals force\cite{Weng-HM}.

\section{RESULTS AND DISCUSSION}
Figure 1d shows representative reflectivity changes $|\Delta R/R|$ at 6, 135 and 300 K, for the cases of n-type semimetal, semiconductor and p-type semimetal, pumped by 3.10 and 1.55 eV optical pulses. In all $|\Delta R/R|$ curves, a wide rising edge immediately appears after the initial photoexcitation ($t_1$), pointing to resolvable relaxation channels in the BCB buildup dynamics. The BCB buildup time $t_2$ is reached at $\sim$ 1 ps (1.55 eV) and $\sim$ 1.5 ps (3.10 eV), respectively. These suggest that tuning the pump to higher photon energy triggers more scattering events in the thermalization process. We also note that $t_2$ agrees with the population time of holes in the valence band and electrons in the conduction band observed by time-resolved photoemission\cite{tr-Band-shift}. In the inset of Fig. 1d, the linear increase of $|\Delta R/R|_{max}$ as a function of pump fluence $F$, implies that the maximum of the reflectivity changes is proportional to the density of photoexcited hot carriers at $t_2$ when the BCB buildup forms.

\begin{figure}                                                                                                                                                     \includegraphics[width=1\columnwidth,angle=0]{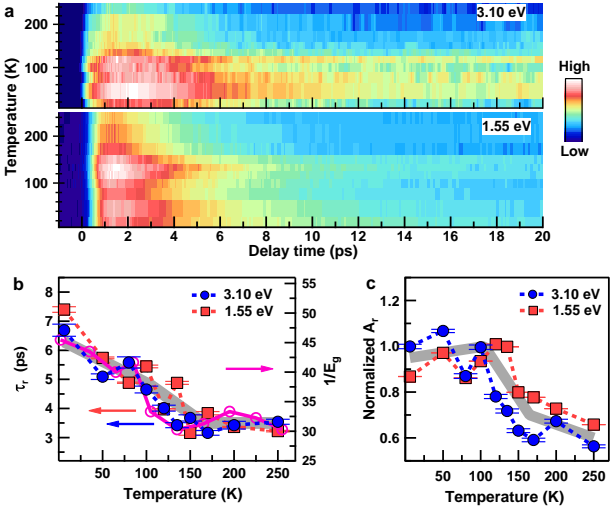}
  \caption{Across-gap relaxation of the BCB buildup after $t_2$. (a) Comprehensive $\Delta R/R$ results with pump fluence $F$ = 18 $\mu$$ J/cm^{2}$ by 3.10 and 1.55 eV. (b) Relaxation time $\tau_{r}$ as a function of temperature, extracted from (a). The error bars are obtained from fits. The pink circles in (b) are the inverse of $T$-dependent gap, confirming $\tau_{r} \varpropto 1/E_g$ (E$_g$ taken from Ref.\cite{Zhou-XJ} with authors' permission). (c) Normalized amplitude $A_{r}$ as a function of temperature. The gray thick curves in (b) and (c) are used to guide the eye for the slope changes at 105 and 165 K. }
\end{figure}

Systematic measurement results are mapped out in Fig. 2a, in which the intensity and relaxation time in $|\Delta R/R|$ significantly increases with temperature decreasing. Let's first focus on the across-gap recombination ($t > t_2$), by using a single exponential function $|\Delta R/R| = A_{r}exp({-t/{\tau_{r}}})+A_0$, where $A_{r}$ and $A_{0}$ correspond to the across-gap recombination and long-lived heating diffusion. In Fig. 2b and 2c, both $\tau_r$ and $A_r$ show two transition temperatures at around 165 and 105 K, indicating that the transient dynamics after $t_2$ is closely related to the gap property. The linear relation between $A_r$  and $F$ in Fig. 4c suggests that $A_r$ is proportional to the number of hot carrier accumulated above the energy gap.

For a gapped system with e-ph coupling, the hot carrier across-gap recombination occurs by the emission of gap-frequency phonons. As phenomenologically described by the Rothwarf-Taylor mode\cite{1967-RTmodel}, gap-frequency phonons/bosons play a dominate role in the reformation dynamics of Cooper pairs in both cuprate\cite{Kabanov-PRB-1999,Pp-SC2,2018-nie-2212} and iron-based\cite{tr-prl-hotcarrier} superconductors and electron-hole pairs in CDW materials\cite{1999-JD-CDWprl}. Smaller gap allows low frequency phonons to be involved in the across-gap recombination process, resulting in a longer lifetime, or even a divergence if the gap is completely closed\cite{2018-nie-2212,1999-JD-CDWprl}. In this case, for simplicity, the relaxation time can be inversely proportional to the gap size $\tau_r \propto 1/E_g$. In Fig. 2b, the $\tau_r(T)$ curves agree well with $1/E_g(T)$, where $E_g(T)$ is the temperature-dependent gap at $E_F$ from ARPES\cite{Zhou-XJ}. In addition, $\tau_r$ is independent of pump photon energy and pump fluence (Fig. 4), confirming that the relaxation bottleneck is dominated by the gap and phonon properties. Simultaneously, from an energy conservation relation $n_{pe}E_g = \epsilon_I$, where $n_{pe} \propto$ A$_r$ is the number of photoexcited hot electrons that relax through the across-gap channel and $\epsilon_I$ is the energy absorption proportional to $F$. Again we obtain a simple relation A$_r \propto$ $1/E_g$. In Fig. 2c, the A$_r(T)$ curve slightly decrease below $\sim$ 105 K and deviates from $1/E_g(T)$. We speculate that here the across-gap probability reduces since few electrons start to occupy the bottom of the conduction band in the n-type semimetal case in equilibrium state.

\begin{figure}                                                                                                                                                 \includegraphics[width=1\columnwidth,angle=0]{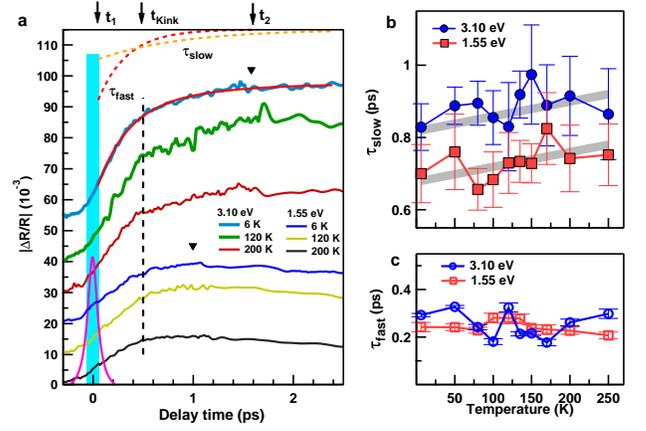}
  \caption{BCB buildup dynamics between $t_1$ and $t_2$.  (a) Reflectivity changes at 6, 135 and 300 K between $t_1$ and $t_2$. The vertical bar presents FWHM of the pump-probe cross-correlation. The red curve is a two-component exponential fit convolved with the width of the cross-correlation. The red and orange dashed lines are single exponential results for decay channels via slow and fast scattering channels. The black dashed line denotes the kink position. The black arrows denote $t_1$, $t_{Kink}$ and $t_2$. The black triangles mark the BCB buildup, the maximum of $|\Delta R/R|$ at $t_2$. (b) and (c) $\tau_{slow}$ and $\tau_{fast}$ as a function of temperature from fits, excited by 3.10 and 1.55 eV respectively. The gray lines are used to guide the increase of $\tau_{slow}$ with temperature increasing.}
\end{figure}

\begin{figure}[t]                                                                                                                                                 \includegraphics[width=1\columnwidth,angle=0]{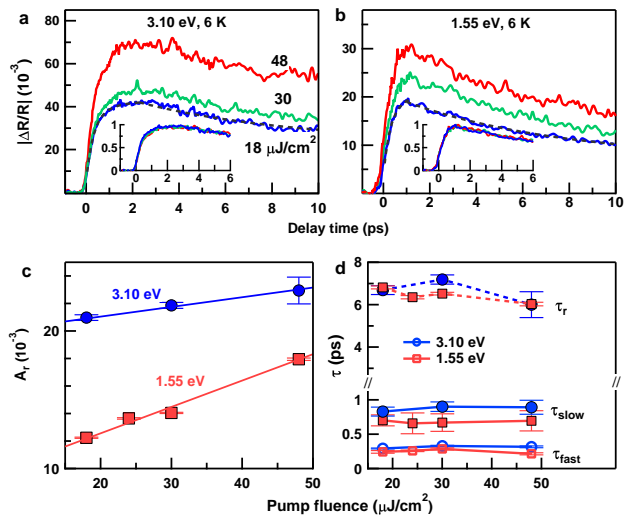}
  \caption{$F$-dependent reflectivity changes. (a) and (b) Representative $F$-dependent differential reflectivity at 6 K excited by 3.10 and 1.55 eV. The rising edge between $t_1$ and $t_2$ is fitted by a two-component exponential function in order to extract $\tau_{e}$ and $\tau_{ph}$, while the relaxation $\tau_{r}$ is obtained by a single exponential fit after $t_2$. The insets are normalized curves, showing good agreement with each other. (c) and (d) $A_{r}$, $\tau_{r}$, $\tau_{ph}$ and $\tau_{e}$ as a function of pump fluence. Error bars are obtained from fits.}
\end{figure}


A deeper inspection of the BCB buildup dynamics, i.e., the zoom-in of the rising edge, is shown in Fig. 3a. As discussed above, the wide rising edge between $t_1$ and $t_2$ refers to the immediate hot carrier redistribution/thermalization after pump light off through e-e and e-ph scattering channels. An universal kink feature in terms of a slope change is clearly observed at a time constant $t_{Kink} \approx$ 500 fs, suggesting coexistence of two scattering channels. We fit the $|\Delta R/R|$ curves from $t_1$ to $t_2$, using a two-component exponential function convolved with a temporal resolution
\bea{}
|\frac{\Delta R}{R}(t)| &= [A_{max} - A_{fast}exp(\frac{-t}{\tau_fast})\nonumber\\
& - A_{slow}exp(\frac{-t}{\tau_{slow}})]*R(\delta t)
\eea
where $R(\delta t)$ is a Gassian curve with FWHM obtained from the cross-correlation measurement.

Figure 3b shows that the slow relaxation, which increases at higher pump photon energy, suggesting more intra/inter-band e-ph scattering events occur in the BCB buildup process. More interestingly, $\tau_{slow}$ increases linearly with lattice temperature increasing and is $F$-independent (Fig. 4d). Obviously the e-ph decay channel here is unable to be described by the effective temperature model, which predicts e-ph scattering rate $\tau_{e-ph} \propto T_e \propto F$\cite{CG-review-2016,2012-JAP-TeTl}, since this model requires high excitation density with electron temperature $T_e$ much higher than lattice temperature. Our results rather agree with the non-equilibrium model, with e-ph scattering written as\cite{2012-JAP-TeTl}
\bea{}
\tau_{e-ph} = \frac{2\pi k_B T_l}{3\hbar\lambda\langle\omega^2\rangle}
\eea
where $\lambda$ is the coupling strength, and $T_l$ is the lattice/sample temperature.

On the other hand, in Fig. 3c and 4d, we find $\tau_{fast}$ = 250 $\pm$ 70 fs, is weakly dependent of temperature, pump photon energy and $F$. One possibility for the origin of $\tau_{fast}$ is the e-e scattering channel, which is more likely a instantaneous process, analogous to that in graphene\cite{graphene2} and semiconductor nanocrystals\cite{2005-Ncrystal-200fs}. Here we consider the impact excitation process in which a low energy electron on the top of the valence band gets excited over the bandgap by absorbing the energy loss from a high-energy hot electron, resulting in an additional electron-hole pair, as depicted in Fig. 1a. The efficiency of the impact excitation relies on the energy and momentum conservation in e-e scattering. In addition, the impact excitation can be very efficient because of high density of carriers as it has been found in graphene\cite{2013-PRB-JCSong}. In ZrTe$_5$, these conditions are possibly to be satisfied owing to its multi-band electronic structure with a bunch of conduction band valleys and high electron density at $\Gamma$. The two-dimensional crystal structure and in-layer carrier density in ZrTe$_5$ ($10^{12} \sim 10^{13} cm^{-2}$\cite{2016-GC-NL}, comparable to that of doped graphene\cite{graphene2}), might enable analogous enhancement of e-e scattering as that in quantum-confined systems. However, we cannot exclude the frequency-dependence of $\Delta R/R$, since the transient dielectric constant varies when hot carriers relax between high-lying conduction bands. Yet more theoretical work and frequency-dependent studies are required to further pin down this behavior.

\section{CONCLUSION}

In summary, we have studied the hot carrier dynamics by ultrafast time-resolved optical spectroscopy. We have clearly evidenced temperature-dependent across-gap recombination dynamics, confirming its gap nature. Further analysis of its electron thermalization dynamics resolves a e-ph scattering channel ruled by the non-equilibrium model. A possible e-e scattering channel is proposed for the fast relaxation, which asks for further investigations. Our results show the importance of many-body effects in ZrTe$_5$ and may further help the understanding of light-matter interaction for its application as photoelectric devices.

\section{ACKNOWLEDGMENTS}
The authors gratefully acknowledge support from the National Natural Science Foundation of China (Grant No. 51275012, 51875006 and 51705009) and
NSAF of China (Grant No. U1530153). HYL thanks the Key Project of Beijing Municipal Natural Science Foundation and Beijing Education Committee¡¯s Science and Technology Plan (Grant No. KZ201810005001).

\end{document}